\documentclass[11pt,english]{article}
\usepackage{cite}
\usepackage{hyperref}
\usepackage{url}
\usepackage{babel}
\usepackage{psfrag,graphicx,float}
\usepackage{amsmath,amssymb,amsthm}
\usepackage[mathscr]{eucal}
\usepackage{microtype}

\newcommand{\CO}{{\mathcal O}}
\newcommand{\CR}{{\mathcal R}}
\newtheorem{cor}{Corollary}

\newtheorem{prop}{Proposition}
\newtheorem{teor}{Theorem}
\def\be{\begin{equation}}
\def\ee{\end{equation}}
\def\ba{\begin{eqnarray}}
\def\ea{\end{eqnarray}}
\def\bs{\begin{subequations}}
\def\es{\end{subequations}}

\newfont{\bbd}{msbm10 scaled\magstep1}

\def\beq{\begin{equation}}
\def\eeq{\end{equation}}
\def\bea{\begin{eqnarray}}
\def\eea{\end{eqnarray}}
\newtheorem{definition}{Definition}[section]

\newcommand{\CA}{{\mathcal A}}
\newcommand{\CQ}{{\mathcal Q}}
\def\ep{\varepsilon}
\def\ri{{\rm{i}}}
\def\endpf{\begin{flushright}$\square$\end{flushright}}

\title {\bf A discrete linearizability test based on multiscale analysis}

\author{{\bf R. Hern\'andez Heredero} \\ Departamento de Matem\'atica Aplicada \\
 Universidad Polit\'ecnica de Madrid,\\
Escuela Universitaria de Ingenier\'ia T\'ecnica de Telecomunicaci\'on,\\
 Campus Sur Ctra.~de Valencia Km.~7, 28031 Madrid, Spain \\
 {\sl E-mail: rafahh@euitt.upm.es} \and {\bf D. Levi} \\ Dipartimento di Ingegneria Elettronica, \\ Universit\`a degli Studi Roma Tre and Sezione INFN, Roma Tre, \\ Via della
Vasca Navale 84, 00146 Roma, Italy \\ {\sl E-mail: levi@roma3.infn.it} \and {\bf C. Scimiterna} \\ Dipartimento di Ingegneria Elettronica, \\ Universit\`a degli Studi Roma Tre and Sezione INFN, Roma Tre, \\ Via della
Vasca Navale 84, 00146 Roma, Italy \\ {\sl E-mail: scimiterna@fis.uniroma3.it} }

\date{\today}
\begin{document}
\maketitle

\begin{abstract}
In this paper we consider the classification of dispersive linearizable partial difference equations defined on a quad-graph by the multiple scale reduction around their harmonic solution.  We show that the $A_1$, $A_2$ and $A_3$ linearizability conditions restrain the number of the parameters which enter into the equation. A subclass of the  equations  which pass the $A_3$ $C-$integrability conditions can be linearized by a M\"obius transformation.

\end{abstract}

%%%%%%%%%%%%%%%%%%%%%%%%%%%%%%%
\section{Introduction}

Calogero in 1991\cite{ca} introduced the notion of S and C integrable equations to characterize those nonlinear partial differential equations which are solvable through an Inverse Scattering Transform or linearizable through a change of variables. Using the multiscale reductive technique he was able to show that the
  Non Linear Schr\"odinger Equation (\emph{NLSE})
\bea
 \mbox{i} \partial_t u = K_2[u]=\partial_{xx} u + \rho_{2} |u|^2 u,\qquad u=u(x,t), \label{e0}
\eea
 appears as a universal equation
governing the evolution of slowly varying packets
of quasi-monochromatic waves in weakly nonlinear
media featuring dispersion. The necessary conditions for the $S$-integrability is that $\rho_2$ is real. If, however, the equation is linearizable then $\rho_2$ must be null, the equation has to be linear or linearizable, as the Eckhaus equation \cite{ce,ls2009}.

By going to a higher order in the expansion, the multiscale techniques have been used  to find new $S-$integrable Partial Differential Equations (PDE's) and   to prove the  integrability of new nonlinear equations\cite{DMS,DP,MK}. Probably the most important example of such nonlinear PDE is  the Degasperis-Procesi equation \cite{dp1}. The application of this test to the case of linearizable equations has not been done yet. However Calogero and collaborators constructed many  interesting linearizable nonlinear evolution equations by applying complicated transformations to  linear equations \cite{ca1,cx}.

In the case of discrete equations it has been shown  \cite{Schoo,Ag,lm,LevHer,levi,lp,HLPS,HLPS2,HLPS3} that a similar situation is also true. One can present the equivalent of the Calogero-Eckhaus theorem stating that a nonlinear dispersive partial difference equation will not be $S-$integrable if its multiscale expansion on analytic functions will not give rise to an integrable~\emph{NLSE} at the lowest order.  Moreover it was shown on examples that a nonlinear partial difference equation will be $C-$integrable if its multiscale expansion on analytic functions will give rise to a linear partial differential equation \cite{ls_sigma}.

As was shown in \cite{HLPS2,HLPS3,levi}, the introduction of multiple scales on a lattice reduces the given discrete equation either to a local nonlinear partial difference equation, by imposing a slow-varying condition, or to a partial differential equation of infinite order when dealing with analytic functions. Here we choose the second alternative because, as was shown in \cite{HLPS2,HLPS3}, only in this case the integrability conditions of the discrete equation are preserved.

In this paper we provide necessary conditions for the linearizability of a class of real difference equations in the variable $u: \mathbb Z^2
\rightarrow \mathbb R$ and its three nearest-neighbors defined on a~$\mathbb{Z}^2$ square-lattice
\beq\label{e}
\CQ (u_{n,m},u_{n+ 1,m},u_{n,m +1}, u_{n + 1,m + 1}; \beta_1, \beta_2,...)=0,
\eeq
where $\beta_i$'s are real parameters. Linearizability conditions will be determined through the  multiple scale perturbative development.

\begin{figure}[htbp]
\begin{center}
\setlength{\unitlength}{0.1em}
\begin{picture}(200,140)(-50,-20)

  \put( 0,  0){\line(1,0){100}}
  \put( 100,  0){\line(-1,0){100}}

  \put( 0,100){\line(1,0){100}}
  \put( 100,100){\line(-1,0){100}}

  \put(  0, 0){\line(0,1){100}}
  \put(  0, 100){\line(0,-1){100}}

  \put(100, 0){\line(0,1){100}}
  \put(100, 100){\line(0,-1){100}}

  %\put(0, 0){\line(1,1){100}}
  %\put(100, 0){\line(-1,1){100}}
   \put(97, -3){$\bullet$}
   \put(-3, -3){$\bullet$}
   \put(-3, 97){$\bullet$}
   \put(97, 97){$\bullet$}
  \put(-32,-13){$u_{n,m}$}
  \put(103,-13){$u_{n+1,m}$}
  \put(103,110){$u_{n+1,m+1}$}
  \put(-32,110){$u_{n,m+1}$}
\end{picture}
\caption{The  $\mathbb{Z}^2$ square-lattice where the equation $\CQ=0$ is defined}
\end{center}
\end{figure}
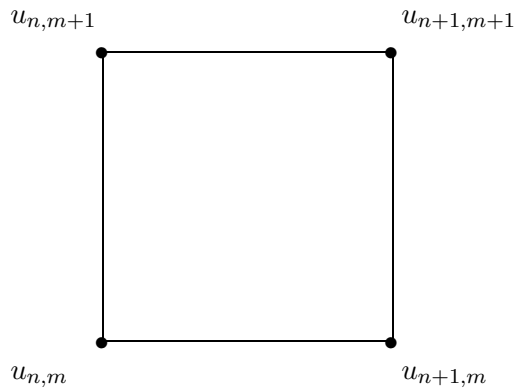

We will suppose that eq.~(\ref{e}) is linear-affine in every variable, and our classification will be carried out up to  the simultaneous M\"obius transformations
\begin{equation}\label{eqMob}
u_{n,m}\mapsto (Au_{n,m}+B)\big/(Cu_{n,m}+D).
\end{equation}

In Section 2 we will briefly discuss  the multiple scale expansion for eq.~(\ref{e}) and present the integrability conditions which ensure  that the given equation is a $C-$integrable equation of $j$ order, with $j=1,2,3$, i.e.~it is such that asymptotically satisfies the $C-$integrability conditions up to third order.
In Section 3  we apply them to the eq.~(\ref{e}) and present a sequence of theorems which give the conditions on the constants $\beta_i$'s under which the system is $A_1$, $A_2$ or $A_3$ integrable. At the end we  present  some conclusive remarks.

%%%%%%%%%%%%%%%%%%%%%%%%%%%%%%%%%%
%%%%%%%%%%%%%%%%%%%%%%%%%%%%%%%%%%
\section{Expansion of real dispersive partial difference equations}
%%%%%%%%%%%%%%%%%%%%%%%%%%%%%%%%%%
%%%%%%%%%%%%%%%%%%%%%%%%%%%%%%%%%%

We now briefly illustrate all the ingredients of the  reductive perturbative technique necessary to treat difference equations, as presented in \cite{HLPS, HLPS2}. We will consider here the multiscale expansion in the case of analytic functions so as to preserve the  linearizability of the given equation.

Introducing multiple lattice scales, under some obvious hypotheses on the $\cal{C}^{(\infty)}$ property of the function $u_{n,m}$ and on the radius of convergence of its Taylor expansion for all the $n$ and $m$ shifts involved in the difference equation (\ref{e}), we can write a series representation of the shifted values of $u_{n,m}$ around the point $(n,m)$. Choosing
\begin{eqnarray}
\nonumber n_i =\ep_{n_{i}} n, \quad  \ep_{n_{i}}\doteq N_{i}\ep^{i},\ \ \ \ \ 1\leq i\leq K_{n},\ \ \ \ \ m_j =\ep_{m_{j}} m, \quad \ep_{m_{j}}\doteq M_{j}\ep^{j},\ \ \ \ \ 1\leq j\leq K_{m},
\end{eqnarray}
where the various constants $N_{i}$, $M_{j}$ and $\ep$ are all real numbers and we will assume $K_{n}=1$ and $K_{m}=K$ (eventually $K=+\infty$),  the total shift operators~$T_n$, $T_m$ can be rewritten in terms of partial shift operators~$\mathcal{T}_n$, $\mathcal{T}_m$ as
\begin{subequations}\label{Istria}
\begin{eqnarray}
&&T_{n}=\mathcal{T}_{n}\mathcal{T}_{n_{1}}^{(\ep_{n_{1}})}=\mathcal{T}_{n}\sum_{j=0}^{+\infty}\ep^j\CA_{n}^{(j)},\ \ \ \ \ \CA_{n}^{(j)}\doteq\frac{N_{1}^{j}} {j!}\partial_{n_{1}}^{j},\label{Istria1}\\
&&T_{m}=\mathcal{T}_{m}\prod_{j=1}^K\mathcal{T}_{m_{j}}^{(\ep_{m_{j}})}=\mathcal{T}_{m}\sum_{j=0}^{+\infty}\ep^j\CA_{m}^{(j)},\\
&&T_{n}T_{m}=\mathcal{T}_{n}\mathcal{T}_{m}\mathcal{T}_{n_{1}}^{(\ep_{n_{1}})}\prod_{j=1}^K\mathcal{T}_{m_{j}}^{(\ep_{m_{j}})}=\mathcal{T}_n\mathcal{T}_m\sum_{j=0}^{+\infty}\ep^j\CA_{n,m}^{(j)},
\end{eqnarray}
\end{subequations}
and in terms of differential operators~$\CA_{m}^{(j)}$, $\CA_{n,m}^{(j)}$. These operators are given by appropriate combinations of $\frac{M_{k}^{j}} {j!}\partial_{m_{k}}^{j}$ (explicit expressions and formulae can be found in~\cite{HLPS}, see \cite{tS} for more details). 
 We can assume for the function $u_{n,m}=u(n,m,n_{1},\left\{m_{j}\right\}_{j=1}^{K},\ep)$ a double expansion in harmonics and in the perturbative parameter $\varepsilon$
\begin{eqnarray}
u_{n,m}=\sum_{\gamma=1}^{+\infty}\sum_{\theta=-\gamma}^{\gamma}\varepsilon^{\gamma}u_{\gamma}^{(\theta)}\left(n_{1},m_{j},j\geq 1\right){\rm e}^{\ri\theta\left(\kappa hn-\omega\sigma m\right)},\label{Catilina}
\end{eqnarray}
with $u_{\gamma}^{(-\theta)}\left(n_{1},m_{j},j\geq 1\right)=\bar u_{\gamma}^{(\theta)}\left(n_{1},m_{j},j\geq 1\right)$ in order to ensure the reality of $u_{n,m}$. Then, inserting the explicit expressions (\ref{Istria}) of the shift operators in terms of the derivatives with respect to the slow variables, eq.~(\ref{e}) turns out to be a PDE of infinite order.
Moreover we  require the functions $u_{\gamma}^{(\theta)}$ to satisfy the asymptotic conditions $\lim_{n_{1}\rightarrow\pm\infty}u_{\gamma}^{(\theta)}=0$, $\forall\gamma$ and $\theta$, and the index $\gamma$ chosen $\geq 1$ in order to let any nonlinear part of eq.~(\ref{e}) to enter as a perturbation in the multiscale expansion.

%%%%%%%%%%%%%%%%%%%%%%%%%%%%%%%%%%%%%%%%%%%%%%%%%%%%%%%%%
\subsection{The orders beyond the Sch\"odinger equation and the C-integrability conditions}
%%%%%%%%%%%%%%%%%%%%%%%%%%%%%%%%%%%%%%%%%%%%%%%%%%%%%%%%%

The multiple scale expansion of a nonlinear partial difference equation  on analytic functions  will thus give rise to continuous PDE's. So a multiple scale linearizability test will require that an equation  is $C-$integrable if its multiple scale expansion will go into the hierarchy of the Schr\"odinger equation. To show this we need to consider  the orders beyond that at which one obtains for the harmonic $u^{\left(1\right)}_{1}$ the Schr\"odinger equation. References~\cite{K1,K2,DMS} contain different approaches to the higher order multiscale expansion of $S-$integrable nonlinear PDE's.

Following \cite{DMS} we can remove all secular terms from the reduced equations order by order and thus, in agreement with Degasperis and Procesi \cite{DP}, we can state the following theorem proved by Michela Procesi \cite{P}:

\begin{teor}
A nonlinear dispersive partial difference equation is $C-$integrable only if its multiscale expansion is given by a uniform asymptotic series such that 

\begin{enumerate}

\item The amplitude $u^{\left(1\right)}_{1}$ evolves at the slow-times $m_{\sigma}$, $\sigma\geq 2$ according to the $\sigma-$th equation of the linear Schr\"odinger hierarchy;
    \bea
    \partial_{m_{\sigma}}u^{\left(1\right)}_{1}=(-i)^{(\sigma-1)} B_{\sigma}\partial_{n_{1}}^{\sigma}u^{\left(1\right)}_{1},\ \ \ \ B_{\sigma}\doteq-\frac{1} {\sigma!}\frac{d^{\sigma}\omega(\kappa)} {d\kappa^{\sigma}},\label{Valentia1}
\eea
where $B_{\sigma}$ are constants.
\item The amplitudes of the higher perturbations of the first harmonic $u^{\left(1\right)}_{j}$, $j\geq 2$ evolve at the slow-times $m_{\sigma}$, $\sigma\geq 2$ according to certain \emph{linear, nonhomogeneous} equations when taking into account the proper \emph{asymptotic boundary conditions}.
\begin{eqnarray}
\partial_{m_{\sigma}}u^{\left(1\right)}_{j }- (-i)^{(\sigma-1)} B_{\sigma}\partial_{n_{1}}^{\sigma}u^{\left(1\right)}_{j}\equiv M_{\sigma} u_j^{(1)}=f_{\sigma}(j),\label{Valentia2}
\end{eqnarray}
$\forall\ j,\ \sigma\geq 2$, where $B_{\sigma}\partial_{n_{1}}^{\sigma}u^{\left(j\right)}_{1}$ is the $\sigma-$th flow in the linear Schr\"odinger hierarchy (\ref{Valentia1}). All the other  $u_{j}^{(\theta)}$, $\theta\geq 2$ are expressed in terms of differential monomials of $u_{\rho}^{(1)}$, $\rho\leq j$.
\end{enumerate}
\end{teor}
If the asymptotic expansion is, besides a necessary condition for integrability, also a sufficient condition, has been only conjectured in \cite{P}. The proof of the theorem takes into account that for a $C-$integrable equation by definition we have a linearizing transformation. The next step is to assume that the function satisfying the linearized equation admits a similar perturbative expansion as in (\ref{Catilina}) and then to perform a multiscale analysis of the linearizing transformation. 

In  eq.~(\ref{Valentia2})  $f_{\sigma}(j)$ is a nonhomogeneous \emph{nonlinear} forcing term and $B_{\sigma}$, $\sigma\geq 1$, introduced in eq.~(\ref{Valentia1}) are complex constants. Eq.~(\ref{Valentia1}) represents a hierarchy of \emph{compatible} evolutions for the  function $u^{\left(1\right)}_{1}$.   It is obvious that the operators $M_{\sigma}$ defined in eq.~(\ref{Valentia2}) commute among themselves.
 Thus, once we fix the index $j\geq 2$ in the set of eqs.~(\ref{Valentia2}), their compatibilities  imply the following conditions
\begin{eqnarray}
M_{\sigma}f_{\sigma'}\left(j\right)=M_{\sigma'}f_{\sigma}\left(j\right),\ \ \ \forall\, \sigma,\sigma'\geq 2,\label{Lavinia}
\end{eqnarray}
where, as $f_{\sigma}\left(j\right)$ and $f_{\sigma'}\left(j\right)$ are functions of the different perturbations of the fundamental harmonic up to degree $j-1$, the time derivatives $\partial_{t_{\sigma}}$, $\partial_{t_{\sigma'}}$ of those harmonics appearing respectively in $M_{\sigma}$ and $M_{\sigma'}$ have to be eliminated using the evolution equations (\ref{Valentia1}, \ref{Valentia2}) up to the index $j-1$. The  commutativity conditions (\ref{Lavinia}) turn out to be {\it an integrability test}.

To  construct the functions $f_{\sigma}^{(j)}$ we define homogeneous spaces of functions of $u_j^{(1)}$, its $n_1$--derivative and its complex conjugate.

\begin{definition}
{\it A differential monomial $\rho\left[u^{\left(1\right)}_{j}\right]$, $j\geq 1$ in the functions $u^{\left(1\right)}_{j}$, its complex conjugate and its $n_1$-derivatives is a monomial of ``gauge'' 1 if it possesses the transformation property
\begin{eqnarray}
\nonumber\rho\left[\tilde u^{\left(1\right)}_{j}\right]= {\rm e}^{\ri\theta}\rho\left[u^{\left(1\right)}_{j}\right],\ \ \ \ \ \tilde u^{\left(1\right)}_{j}\doteq {\rm e}^{\ri\theta}u^{\left(1\right)}_{j};
\end{eqnarray}}
\end{definition}

\begin{definition}\label{FAVL1}
{\it A finite dimensional vector space $\mathcal P_{\nu}$, $\nu \geq 2$ is the set of all differential polynomials in the functions $u^{\left(1\right)}_{j}$, $j\geq 1$, their complex conjugates and their $n_1$-derivatives of order $\nu$ in $\ep$ and gauge 1 such that
\begin{eqnarray}
\mbox{order}\left(\partial_{n_1}^{\mu}u^{\left(1\right)}_{j}\right)=\mbox{order}\left(\partial_{n_1}^{\mu}\bar u^{\left(1\right)}_{j}\right)=\mu+j,\ \ \ \mu\geq 0;\nonumber
\end{eqnarray}}
\end{definition}

\begin{definition}\label{FAVL2}
{\it $\mathcal P_{\nu}(\mu)$, $\mu\geq 1$ and $\nu\geq 2$ is the subspace of $\mathcal P_{\nu}$ whose elements are differential polynomials in the functions $u^{\left(1\right)}_{j}$s, their complex conjugates and their $n_1$-derivatives of order $\nu$ in $\ep$ and gauge 1 for $1\leq j\leq \mu$.}
\end{definition}

 The basis monomials of the spaces $\mathcal P_{\nu}(\mu)$ can be found, for example, in \cite{tS}.

The functions $f_{\sigma}(j)$ belong to a well defined finite dimensional space $\mathcal P_{\nu}(\mu)$ and if 
 the relations (\ref{Lavinia}) are satisfied up to the index $k$, $k\geq 2$, we say that our equation is asymptotically $C-$integrable of degree $k$ or $A_{k}$ $C-$integrable.

\subsection{$C-$integrability conditions for the Schr\"odinger hierarchy}

In this subsection we present the conditions which must be satisfied for the asymptotic $C-$integrability of order $k$ or the $A_{k}$ $C-$integrability conditions with $k=1,2,3$. To simplify the notation, we will use for $u^{\left(1\right)}_{j}$ the concise form $u(j)$.

 The {\bf  $A_{1}$ $C-$integrability condition} is given by the absence of the coefficient $\rho_{2}$ of the nonlinear term in the~\emph{NLSE}.
 % It is obtained commuting the~\emph{NLSE} flux $K_{2}[u]$ with the higher order flux $B\left[u_{\xi\xi\xi}+\tau |u|^2u_{\xi}+\mu u^2\bar u_{\xi}\right]$ with $\tau$ and $\mu$ constants. This commutativity condition gives, if $\rho_{2} =0$, $\tau=\mu=0$  and no conditions on $B$ and $\rho_1$ although they will always turn out to be real.

 The {\bf $A_{2}$ $C-$integrability condition} is obtained by choosing $j=2$ in the compatibility conditions (\ref{Lavinia}) with $\sigma=2$ and $\sigma'=3$:
\begin{eqnarray}
M_{2}f_{3}\left(j\right)=M_{3}f_{2}\left(j\right),\label{Turno}
\end{eqnarray}
where $M_2 = \partial_{m_2} +i B_2 \partial^2_{n_1}$ and $M_3 = \partial_{m_3} + B_3 \partial^3_{n_1}$.

In this case $f_{2}(2)\in\mathcal P_{4}(1)$ and $f_{3}(2)\in\mathcal P_{5}(1)$ with ${\rm dim}(\mathcal P_{4}(1))=2$ and ${\rm dim}(\mathcal P_{5}(1))=5$, so that $f_{2}(2)$ and $f_{3}(2)$ will be respectively identified by 2 and 5 complex constants
\begin{subequations}
\begin{eqnarray}
&& f_{2}(2)\doteq au_{n_1}(1)|u(1)|^2+b\bar u_{n_1}(1)u(1)^2,\label{Abruzzo1}\\
&& f_{3}(2)\doteq\alpha |u(1)|^4u(1)+\beta |u_{n_1}(1)|^2u(1)+\gamma u_{n_1}(1)^2\bar u(1)+\label{Abruzzo2}\\
&&\ \ \ \ \ \ \ \ +\delta\bar u_{n_1 n_1}(1)u(1)^2+\epsilon |u(1)|^2u_{n_1 n_1}(1).\nonumber
\end{eqnarray}
\end{subequations}
In this way, eliminating from eq.~(\ref{Turno}) the derivatives of $u(1)$ with respect to the slow-times $m_{2}$ and $m_{3}$ using the evolutions (\ref{Valentia1}) with $\sigma=2$, 3 and equating term by term, we obtain that the  $A_{2}$ $C-$integrability conditions give no constraint for the two coefficients $a$ and $b$. As a consequence one can say that the $A_{1}$ $C-$integrability condition $\rho_{2}=0$ automatically implies $A_{2}$ $C-$integrability, or that it indeed represents an $A_{2}$ $C-$integrability condition. The expression of $\alpha$, $\beta$, $\gamma$, $\delta$, $\epsilon$ in terms of $a$ and $b$ are:
\begin{eqnarray}
\alpha= 0,\ \ \ \beta=-\frac{3\ri B_3b} {B_2},\ \ \ \gamma=-\frac{3\ri B_3a} {2B_2},\ \ \ \delta=0,\ \ \ \epsilon=\gamma.\label{Molise}
\end{eqnarray}
where, for convenience, from now on we will write $B$ for $B_3$.

 The {\bf $A_{3}$ $C-$integrability conditions} are derived in a similar way setting $j=3$ in eq.~(\ref{Turno}). In this case we have that $f_{2}(3)\in\mathcal P_{5}(2)$ and $f_{3}(3)\in\mathcal P_{6}(2)$ with ${\rm dim}(\mathcal P_{5}(2))=12$ and ${\rm dim}(\mathcal P_{6}(2))=26$, so that $f_{2}(3)$ and $f_{3}(3)$ will be respectively identified by 12 and 26 complex constants
\begin{subequations} \label{f23}
\begin{align}
\begin{split}
f_{2}(3)&\doteq\tau_{1}|u(1)|^4u(1)+\tau_{2}|u_{n_1}(1)|^2u(1)+\tau_{3}|u(1)|^2u_{n_1 n_1}(1)+\tau_{4}\bar u_{n_1 n_1}(1)u(1)^2\\
&+\tau_{5}u_{n_1}(1)^2\bar u(1)+\tau_{6}u_{n_1}(2)|u(1)|^2+\tau_{7}\bar u_{n_1}(2)u(1)^2+\tau_{8}u(2)^2\bar u(1)+\tau_{9}|u(2)|^2u(1)
\\
&+\tau_{10}u(2)u_{n_1}(1)\bar u(1)+\tau_{11}u(2)\bar u_{n_1}(1)u(1)+\tau_{12}\bar u(2)u_{n_1}(1)u(1),
\end{split}
\label{Lazio4}\\
\begin{split}
f_{3}(3)&\doteq\gamma_{1}|u(1)|^4u_{n_1}(1)+\gamma_{2}|u(1)|^2u(1)^2\bar u_{n_1}(1)+\gamma_{3}|u(1)|^2u_{n_1 n_1 n_1}(1)
\\
&+\gamma_{4}u(1)^2\bar u_{n_1 n_1 n_1}(1)+\gamma_{5}|u_{n_1}(1)|^2u_{n_1}(1)+\gamma_{6}\bar u_{n_1 n_1}(1)u_{n_1}(1)u(1)\\
&+\gamma_{7}u_{n_1 n_1}(1)\bar u_{n_1}(1)u(1)
+\gamma_{8}u_{n_1 n_1}(1)u_{n_1}(1)\bar u(1)+\gamma_{9}|u(1)|^4u(2)
\\
&+\gamma_{10}|u(1)|^2u(1)^2\bar u(2)+\gamma_{11}\bar u_{n_1}(1)u(2)^2+\gamma_{12}u_{n_1}(1)|u(2)|^2 +\gamma_{13}|u_{n_1}(1)|^2u(2)
\\
&+\gamma_{14}|u(2)|^2u(2)+\gamma_{15}u_{n_1}(1)^{2}\bar u(2)+\gamma_{16}|u(1)|^2u_{n_1 n_1}(2)+\gamma_{17}u(1)^2\bar u_{n_1 n_1}(2)
\\
&+\gamma_{18}u(2)\bar u_{n_1 n_1}(1)u(1)+\gamma_{19}u(2)u_{n_1 n_1}(1)\bar u(1) +\gamma_{20}\bar u(2)u_{n_1 n_1}(1)u(1)
\\
&+\gamma_{21}u(2)u_{n_1}(2)\bar u(1)+\gamma_{22}\bar u(2)u_{n_1}(2)u(1)+\gamma_{23}u_{n_1i}(2)u_{n_1}(1)\bar u(1)
\\
&+\gamma_{24}u_{n_1}(2)\bar u_{n_1}(1)u(1)+\gamma_{25}\bar u_{n_1i}(2)u_{n_1}(1)u(1)+\gamma_{26}\bar u_{n_1}(2)u(2)u(1).
\end{split}\label{Lazio5}
\end{align}
\end{subequations}
Let us eliminate from eq.~(\ref{Turno}) with $j=3$ the derivatives of $u(1)$ with respect to the slow-times $m_{2}$ and $m_{3}$ using the evolutions (\ref{Valentia1}) respectively with $\sigma=2$, 3 and the same derivatives of $u(2)$ using the evolutions (\ref{Valentia2}) with $\sigma=2$, 3. Equating the remaining terms term by term, the $A_{3}$ $C-$integrability conditions turn out to be:
\begin{equation} \label{Aurunci}
\begin{gathered}
\tau_{1}=-\frac{\ri} {4B_2}\left[b\left(\tau_{11}-2\tau_{6}\right)+\bar a\tau_{7}\right],\qquad \bar b\tau_{7}=\frac{1} {2}\left(b-a\right)\left(\tau_{11}+\tau_{10}-\tau_{6}\right)+\bar a\tau_{7},\\
a\tau_{8}=b\tau_{8}=0,\qquad a\tau_{9}=b\tau_{9}=0,\qquad \bar a\tau_{12}=a\left(\tau_{10}-\tau_{11}\right)+b\tau_{6}+\bar a\tau_{7},\\
\left(\bar b-\bar a\right)\tau_{12}=\left(b-a\right)\tau_{10}.\end{gathered}
\end{equation}
Sometimes $a$ and $b$ turn out to be both real. In this case the conditions given in eqs.~({\ref{Aurunci}}) become:
\begin{equation}\label{Aurunci2}
\begin{gathered}
R_{1}=\frac{1} {4B_2}\left[b\left(I_{11}-2I_{6}\right)+aI_{7}\right],\qquad I_{1}=-\frac{1} {4B_2}\left[b\left(R_{11}-2R_{6}\right)+aR_{7}\right],
\\
\left(b-a\right)\left(R_{11}+R_{10}-R_{6}-2R_{7}\right)=0,\qquad \left(b-a\right)\left(I_{11}+I_{10}-I_{6}-2I_{7}\right)=0,
\\
\left(b-a\right)R_{8}=0,\qquad \left(b-a\right)I_{8}=0,\qquad \left(b-a\right)R_{9}=0,\qquad \left(b-a\right)I_{9}=0,
\\
a\left(R_{12}+R_{11}-R_{10}-R_{7}\right)=bR_{6},\qquad a\left(I_{12}+I_{11}-I_{10}-I_{7}\right)=bI_{6},
\\
\left(b-a\right)\left(R_{12}-R_{10}\right)=0,\qquad \left(b-a\right)\left(I_{12}-I_{10}\right)=0,
\end{gathered}
\end{equation}
where $\tau_{j}=R_{j}+\ri I_{j}$ for $j=1,\ldots, 12$. The expressions of the $\gamma_{j}$ as functions of the $\tau_{i}$ are:

\begin{gather*}
\gamma_{1}=\frac{3B_3} {4B_2^2}\left(a\tau_{6}-4\ri B_2\tau_{1}+\bar b\tau_{12}\right),\quad \gamma_{2}=\frac{3B_3} {4B_2^2}\left(b\tau_{6}+\bar a\tau_{7}\right),
\\
\gamma_{3}=-\frac{3\ri B_3\tau_{3}} {2B_2},\quad \gamma_{4}=0,\quad \gamma_{5}=-\frac{3\ri B_3\tau_{2}} {2B_2},
\\
\gamma_{6}=-\frac{3\ri B_3\tau_{4}} {B_2},\quad \gamma_{7}=\gamma_{5},\quad \gamma_{8}=\gamma_{3}-\frac{3\ri B_3\tau_{5}} {B_2},\quad \gamma_{9}=0,\quad \gamma_{10}=0,
\\
\gamma_{11}=0,\quad \gamma_{12}=-\frac{3\ri B_3\tau_{9}} {2B_2},\quad \gamma_{13}=-\frac{3\ri B_3\tau_{11}} {2B_2},\quad \gamma_{14}=0,\quad \gamma_{15}=-\frac{3\ri B_3\tau_{12}} {2B_2},
\\
\gamma_{16}=-\frac{3\ri B_3\tau_{6}} {2B_2},\quad \gamma_{17}=\gamma_{18}=0,\quad \gamma_{19}=-\frac{3\ri B_3\tau_{10}} {2B_2},\quad \gamma_{20}=\gamma_{15},\quad \gamma_{21}=-\frac{3\ri B_3\tau_{8}} {B_2},
\\
\gamma_{22}=\gamma_{12},\quad \gamma_{23}=\gamma_{16}+\gamma_{19},\quad \gamma_{24}=\gamma_{13},\quad \gamma_{25}=-\frac{3\ri B_3\tau_{7}} {B_2},\quad \gamma_{26}=0.
\end{gather*}
The conditions given in eqs.~({\ref{Aurunci}}, {\ref{Aurunci2}}) appear to be new. Their importance resides in the fact that a $C-$integrable equation must satisfy those conditions.

%%%%%%%%%%%%%%%%%%%%%%%%%%%%%%%
%%%%%%%%%%%%%%%%%%%%%%%%%%%%%%%
\section{Dispersive affine-linear equations on the square lattice} \label{s2}
%%%%%%%%%%%%%%%%%%%%%%%%%%%%%%%
%%%%%%%%%%%%%%%%%%%%%%%%%%%%%%%

In this Section we derive the
necessary conditions for the linearizability of
a  dispersive and homogeneous affine-linear equation defined on the square lattice and belonging to the class (\ref{e}).
The most general multilinear  equation of the class (\ref{e}) has at most quartic nonlinearity. Let us  introduce into its linear part the solution $u_{n,m}= K^n \Omega^m$ where $K={\rm e}^{\ri\kappa}$ and $\Omega={\rm e}^{-\ri\omega(\kappa)}$. We get that this equation is dispersive if homogeneous and  written as
\begin{equation}\label{q4}
\begin{aligned}
\CQ_\pm&=  a_1 (u_{n,m} \pm u_{n+1,m+1}) + a_2 (u_{n+1,m} \pm u_{n,m+1}) \\
&{}+  (\alpha_1-\alpha_2) \, u_{n,m}u_{n+1,m} +   (\alpha_1+\alpha_2)\,
 u_{n,m+1}u_{n+1,m+1} 
\\
&{}+  (\beta_1-\beta_2)\, u_{n,m}u_{n,m+1} + (\beta_1+\beta_2)\,
u_{n+1,m}u_{n+1,m+1} 
\\
&{}+ \gamma_1 u_{n,m}u_{n+1,m+1} + \gamma_2 u_{n+1,m}u_{n,m+1} 
\\
&{}+ (\xi_1-\xi_3)\, u_{n,m}u_{n+1,m}u_{n,m+1} + (\xi_1+\xi_3)\,
u_{n,m}u_{n+1,m}u_{n+1,m+1} 
\\
&{}+ (\xi_2-\xi_4)\, u_{n+1,m}u_{n,m+1}u_{n+1,m+1} + (\xi_2+\xi_4)\,
u_{n,m}u_{n,m+1}u_{n+1,m+1} 
\\
&{}+  \zeta u_{n,m} u_{n+ 1,m} u_{n,m + 1} u_{n + 1,m + 1}=0,
\end{aligned}
\end{equation}
where $a_1,a_2 \in \mathbb R\setminus \{0\}$, $|a_1| \neq |a_2|$, are the coefficients appearing in the
linear part while  $\alpha_1,\alpha_2,\beta_1,\beta_2,$ $\gamma_1,\gamma_2,$
$\xi_1,...,\xi_4, \zeta$ are eleven real parameters to be determined by using
the multiscale procedure described in Section 2.
The linear  dispersion relation is given by
\beq
\omega(\kappa) = {\rm{arctan}} \left[ \frac{(a_1^2-a_2^2) \sin \kappa}{
(a_1^2+a_2^2) \cos \kappa + 2 a_1 a_2} \right]. \label{disp}
\eeq
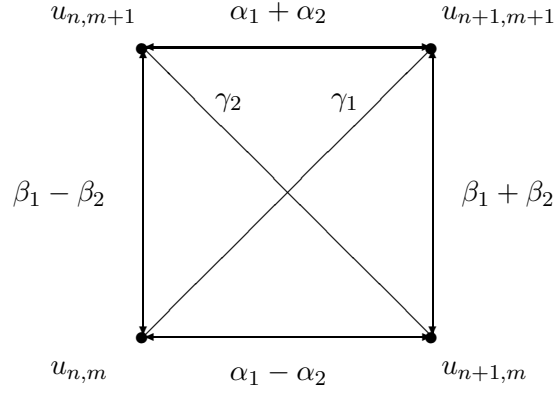
\begin{figure}[htbp]
\begin{center}
\setlength{\unitlength}{0.1em}
\begin{picture}(200,140)(-50,-20)

  \put( 0,  0){\vector(1,0){100}}
  \put( 100,  0){\vector(-1,0){100}}

  \put( 0,100){\vector(1,0){100}}
  \put( 100,100){\vector(-1,0){100}}

  \put(  0, 0){\vector(0,1){100}}
  \put(  0, 100){\vector(0,-1){100}}

  \put(100, 0){\vector(0,1){100}}
  \put(100, 100){\vector(0,-1){100}}

  \put(0, 0){\line(1,1){100}}
  \put(100, 0){\line(-1,1){100}}
   \put(97, -3){$\bullet$}
   \put(-3, -3){$\bullet$}
   \put(-3, 97){$\bullet$}
   \put(97, 97){$\bullet$}
  \put(-32,-13){$u_{n,m}$}
    \put(-45,47){$\beta_1-\beta_2$}
    \put(110,47){$\beta_1+\beta_2$}
     \put(30,-15){$\alpha_1-\alpha_2$}
     \put(30,110){$\alpha_1+\alpha_2$}
     \put(65,80){$\gamma_1$}
     \put(25,80){$\gamma_2$}
  \put(103,-13){$u_{n+1,m}$}
  \put(103,110){$u_{n+1,m+1}$}
  \put(-32,110){$u_{n,m+1}$}
\end{picture}
\caption{Representation of the quadratic nonlinearities of $\CQ_\pm$}
\end{center}
\end{figure}

Let us  look for those transformations which leave the class of equations $\CQ_\pm$ invariant. They  will give the equivalence conditions  for our classification. It is well known that polynomial  equations are invariant under simultaneous M\"obius transformations~$u_{n,m}\mapsto (Au_{n,m}+B)\big/(Cu_{n,m}+D)$. However,  our class of equations is homogeneous with a restriction on the  coefficients of the linear part. After a M\"obius transformation a constant term
\begin{equation}\label{eqConst}
a_0=
B^4\zeta+
2 B^3D
   \left(\xi _1+\xi
   _2\right)+ B^2 D^2
   \left[\gamma_1+\gamma_2+2
   \left(\alpha _1+\beta
   _1\right)\right]+2BD^3
   \left(a_1+a_2\right)
\end{equation}
will appear, and thus, if we do not want to restrict the coefficients of the equation,  we have to set $B=0$  to have $a_0=0$. Under the M\"obius transformation with $B=0$ the coefficients of $\CQ_\pm$ become
\begin{gather}
a_1\mapsto D^3a_1,\quad a_2\mapsto D^3a_2,\quad \alpha_1\mapsto D^2 \left[\alpha _1+C\left(a_1+a_2\right)
   \right],\quad \alpha_2\mapsto D^2\alpha_2,
\\
\beta_1\mapsto D^2 \left[\beta _1+C\left(a_1+a_2\right)
   \right],\quad \beta_2\mapsto D^2\beta_2,
\\
 \gamma_1\mapsto D^2 \left(\gamma_1+2C a_1 \right),\quad \gamma_2\mapsto D^2 \left(\gamma_2+2C a_2 \right),
\\
\xi_1\mapsto D \xi _1+\tfrac{1}{2} C D \left[3C
   \left(a_1+a_2\right)
   +\gamma_1+\gamma_2+2 \left(\alpha
   _1-\alpha _2+\beta
   _1\right)\right],
\\
\xi_2\mapsto D \xi _2+\tfrac{1}{2} C D \left[3C
   \left(a_1+a_2\right)
   +\gamma_1+\gamma_2+2 \left(\alpha
   _1+\alpha _2+\beta
   _1\right)\right],
\\
\xi_3\mapsto D \xi _3+\tfrac{1}{2} C D \left[C(a_1
   -a_2)+\gamma_1-\gamma_2+2 \beta
   _2\right],
\\
\xi_4\mapsto D \xi _4+\tfrac{1}{2} C D \left[C(a_1
   -a_2)+\gamma_1-\gamma_2-2 \beta
   _2\right],
\\
\zeta\mapsto \zeta+C^2 \left[2C
   \left(a_1+a_2\right)
   +\gamma_1+\gamma_2+2 \left(\alpha
   _1+\beta _1\right)\right]+2C
   \left(\xi _1+\xi
   _2\right).\label{eqtrzeta}
\end{gather}
So our equivalence transformation with respect to which we will be classifying the equation $\CQ_ \pm$ is a restricted M\"obius transformation of the form~$u_{n,m}\mapsto u_{n,m}/(Cu_{n,m}+D)$. 

In this article we limit ourselves to consider the case of $\CQ_+$. The case $\CQ_-$ will be dealt with in a subsequent publication.

Imposing that $\rho_2=0$ we get he following Proposition:
\begin{prop} \label{the2}
The lowest order necessary conditions for the linearizability of
  equations $\CQ_+$ give 6 different classes of equations characterized by different non--superimposed ranges of values of the coefficients of the equation $\CQ_+$. They are:
\begin{itemize}
\item Case L1:
 \beq \label{L1}
\left\{ \begin{array}{l}
\alpha_2= \beta_2=0 ,\quad \alpha_1 = \beta_1,\quad\gamma_1+\gamma_2=2\beta_1,  \\
\xi_1 = \xi_2=\frac{\left(a_1+a_2\right){}^2
   \gamma_1 \gamma_2+\left(3 a_1-2 a_2\right)a_2 \gamma_1 \beta
   _1-a_1 \left(2 a_1-3 a_2\right) \gamma_2
   \beta _1}{4 a_1a_2 \left(a_1+a_2\right)},   \\
\xi_3 = \xi_4=
   \frac{-\left(a_1-a_2\right) \left(a_1+a_2\right){}^2 \gamma_1 \gamma_2-a_2 \left(-a_1^2-5 a_2 a_1+2 a_2^2\right) \gamma_1 \beta _1+a_1
   \left(2 a_1^2-5 a_2 a_1-a_2^2\right) \gamma_2 \beta
   _1}{4 a_1 a_2
   \left(a_1+a_2\right){}^2}.
\end{array} \right.
\eeq
\item Case L2:
 \beq \label{L2}
\left\{ \begin{array}{l}
\alpha_2= \beta_2=0 ,\quad \alpha_1 = \beta_1,\quad
\left(3 a_1-2
   a_2\right) a_2^2\gamma_1+a_1^2 \left(2
   a_1-3 a_2\right)
   \gamma_2= 4 a_1
   \left(a_1-a_2\right) a_2
   \beta _1,  \\
\xi_1 = \xi_2=\frac{
\left(a_1+a_2\right)(a_2^2
    \gamma_1^2-a_1^2
   \gamma_2^2)+2 a_2 \left(a_1^2-2 a_2^2\right) \gamma_1 \beta _1+2 a_1 \left(2
   a_1^2-a_2^2\right) \gamma_2 \beta _1-6 a_1 \left(a_1-a_2\right) a_2 \beta _1^2}
   {4 a_1 a_2 \left(a_1^2-a_2^2\right)},   \\
\xi_3 = \xi_4=
   \frac{2 a_1 a_2 \left(a_1+a_2\right) \left(\gamma_1-\gamma_2\right) \beta
   _1+\left(a_1-a_2\right) \left(a_2 \gamma_1-a_1
   \gamma_2\right){}^2+2 a_1 a_2 \left(a_2-a_1\right) \beta _1^2
}{4 a_1 a_2 \left(a_1+a_2\right)^2}.
\end{array} \right.
\eeq
\item Case L3:
 \beq \label{L3}
\left\{ \begin{array}{l}
\alpha_2= \beta_2=0 ,\quad a_1=2a_2,\quad
\gamma_1=\frac{2}{3}
   \left(\alpha _1+\beta
   _1\right),\quad
\gamma_2=\frac{1}{3}
   \left(\alpha _1+\beta
   _1\right),  \\
\xi_1 = \xi_2=-\frac{-5 \alpha _1
   \beta _1+\alpha _1^2+\beta
   _1^2}{6 a_2},\quad
\xi_3 = \xi_4=-\frac{\left(\alpha _1-2
   \beta _1\right) \left(2
   \alpha _1-\beta
   _1\right)}{18 a_2}.
\end{array} \right.
\eeq
\item Case L4:
 \beq \label{L4}
\left\{ \begin{array}{l}
\alpha_2= \beta_2=0 ,\quad 2a_1=a_2,\quad
\gamma_1=\frac{1}{3}
   \left(\alpha _1+\beta
   _1\right),\quad
\gamma_2=\frac{2}{3}
   \left(\alpha _1+\beta
   _1\right),  \\
\xi_1 = \xi_2=-\frac{-5 \alpha _1
   \beta _1+\alpha _1^2+\beta
   _1^2}{6 a_1},\quad \xi_3 = \xi_4=\frac{\left(\alpha _1-2
   \beta _1\right) \left(2
   \alpha _1-\beta
   _1\right)}{18 a_1}.
\end{array} \right.
\eeq
\item Case L5:
 \beq \label{L5}
\left\{ \begin{array}{l}
\alpha_2= \beta_2 ,\quad \alpha_1= \beta_1 ,\quad 2a_1=a_2,\quad
\gamma_1=\frac{2
   \beta _1}{3},\quad \gamma_2=\frac{4
   \beta _1}{3},  \\
\begin{aligned}
\xi_1 &= \tfrac{3
   \beta _1^2-2 \beta _2 \beta
   _1+\beta _2^2}{6 a_1},\quad&
\xi_2&=\tfrac{3 \beta _1^2+2
   \beta _2 \beta _1+\beta
   _2^2}{6 a_1},   \\
\xi_3 &= -\tfrac{\beta _1^2-6 \beta
   _2 \beta _1+7 \beta
   _2^2}{18 a_1},\quad&
\xi_4&=-\tfrac{\beta _1^2+6 \beta
   _2 \beta _1+7 \beta
   _2^2}{18 a_1}.
\end{aligned}
\end{array} \right.
\eeq
\item Case L6:
 \beq \label{L6}
\left\{ \begin{array}{l}
\alpha_2= - \beta_2 ,\quad \alpha_1= \beta_1 ,\quad a_1=2a_2,\quad
\gamma_1=\frac{4 \beta _1}{3},\quad
\gamma_2=\frac{2 \beta _1}{3},  \\
\begin{aligned}
\xi_1 &= \tfrac{3
   \beta _1^2+2 \beta _2 \beta
   _1+\beta _2^2}{6 a_2}&\quad
\xi_2&=\tfrac{3 \beta _1^2-2
   \beta _2 \beta _1+\beta
   _2^2}{6 a_2},   \\
\xi_3 &= \tfrac{\beta _1^2+6 \beta _2
   \beta _1+7 \beta _2^2}{18
   a_2} &\quad
\xi_4&=\tfrac{\beta
   _1^2-6 \beta _2 \beta _1+7
   \beta _2^2}{18 a_2}.
\end{aligned}
\end{array} \right.
\eeq
\end{itemize}
The corresponding six subclasses of equations are invariant under restricted M\"obius transformations of the form $u_n\mapsto u_n/(Cu_n+D)$.
\end{prop}

{\underline{Proof:}} Following the procedure described in Section 2 we
expand the fields appearing in equation $\CQ_+$ using the definitions
(\ref{Catilina}, \ref{Istria}).
The lowest order necessary conditions for the integrability of $\CQ_+$ are obtained
at the order
$\ep^3$ of the multiple scale expansion. At this order we get
the $m_2$-evolution equation for the harmonic $u_1^{(1)}$, that is in general  a nonlinear Shr\"odinger
equation of the form
\beq
\ri  \partial_{m_2} u_1^{(1)} -B_2 \partial_{n_2}^2 u_1^{(1)}-
 \rho_2 u_1^{(1)} |u_1^{(1)}|^2=0, \quad \mbox{where},\quad n_2 = n_1 - \frac{d \omega}{d \kappa} m_1, \label{nls}
\eeq
where the coefficients $B_2$ and $\rho_2$ will depend on the parameters of
the equation $\CQ_+$ and on the wave parameters $\kappa$ and $\omega$, with
$\omega$ expressed in terms of $\kappa$ through the dispersion relation (\ref{disp}).

Let us give just a sketch of the construction of eq.~(\ref{nls}), omitting all
intermediate formulas.
At $\CO(\ep)$ we get:
\begin{itemize}
\item for $\theta=1$ a linear equation which is identically satisfied by the
  dispersion relation (\ref{disp}).
\item for $\theta=0$ a linear equation whose solution is $u_1^{(0)}=0$.
\end{itemize}
At $\CO(\ep^2)$, taking into account the dispersion relation (\ref{disp}),
we get:
\begin{itemize}
\item for $\theta=2$ an algebraic relation between $u_2^{(2)}$ and $u_1^{(1)}$.
\item for $\theta=1$ a linear wave equation for $u_1^{(1)}$, whose solution is
  given by $u_1^{(1)}(n_1, m_1, m_2)=u_1^{(1)}(n_2, m_2)$, with
$n_2$ given by eq.~(\ref{nls}).
\item for $\theta=0$ an algebraic relation between $u_1^{(1)}$ and $u_2^{(0)}$.
\end{itemize}
Let us stress here  that at $\CO(\ep^2)$ we find that  all the harmonics will  depend
on the slow-variables $n_1$ and $m_1$ through $n_2$.

At $\CO(\ep^3)$, for $\theta=1$, by using  the previous results, one gets the \emph{NLSE} (\ref{nls}) with
$$
B_2 = -\frac{a_1 a_2 (a_1^2-a_2^2) \sin \kappa }{ (a_1^2+a_2^2+2 a_1 a_2
  \cos \kappa)^2}, \qquad
\rho_2 = \CR_1 + \ri \CR_2, %\label{r2}
$$
where
\bea
&& \CR_1=\frac{\sin \kappa \left[ \CR_1^{(0)} + \CR_1^{(1)} \cos \kappa
+ \CR_1^{(2)} \cos^2 \kappa+ \CR_1^{(3)} \cos^3 \kappa +\CR_1^{(4)} \cos^4 \kappa  \right]}
{(a_1+a_2)(a_1^2+a_2^2+2 a_1 a_2 \cos \kappa)^2 \left[(a_1-a_2)^2 + 2 a_1 a_2 \cos
  \kappa (1+ \cos \kappa) \right]}, \label{r21} \\
&& \CR_2=\frac{\CR_2^{(0)} + \CR_2^{(1)} \cos \kappa
+ \CR_2^{(2)} \cos^2 \kappa+ \CR_2^{(3)} \cos^3 \kappa +\CR_2^{(4)} \cos^4
\kappa
+\CR_2^{(5)} \cos^5 \kappa}
{(a_1+a_2)(a_1^2+a_2^2+2 a_1 a_2 \cos \kappa)^2 \left[(a_1-a_2)^2 + 2 a_1 a_2 \cos
  \kappa (1+ \cos \kappa) \right]}. \label{r22}
\eea
Here the coefficients $\CR_1^{(i)}$, $0 \leq i \leq 4$, and $\CR_2^{(i)}$, $0
\leq i \leq 5$, are polynomials
depending on the coefficients $a_1,a_2 ,\alpha_1,\alpha_2,\beta_1,\beta_2,$ $\gamma_1,\gamma_2,$
$\xi_1,...,\xi_4$ and their expressions are cumbersome, so that we omit
them.

Note that $B_2$ is a real coefficient depending only on the parameters
of the linear part of $\CQ_+$, while $\rho_2$ is a complex one. Hence
the linearizability of the~\emph{NLS} equation (\ref{nls}) is equivalent to the request
$\rho_2 =0 \; \forall \, \kappa$, that is
\beq
\CR_j^{(i)}=0, \qquad 0\leq i \leq 5, \, j=1,2. \label{sys}
\eeq
 Eq.~(\ref{sys}) is a nonlinear algebraic system of eleven equations in twelve
unknowns, the coefficient~$\zeta$ not appearing at this order of the multiple scale expansion.
By solving it with the help of the computer algebra software Mathematica one gets the solutions (\ref{L1}--\ref{L6}).
These solutions are computed taking into account that
$a_1,a_2 \in \mathbb R\setminus \{0\}$ with $|a_1| \neq |a_2|$. 
Let us point out that the equations have been solved using only rational algebra, avoiding the unreliability of computer algebra calculations when using irrational functions.

One first solves the six equations corresponding to $\CR_2^{(i)}=0$,  $0\leq i \leq 5$.
Two of them can be solved for $\xi_1$
and $\xi_3$ in (rational) terms of the remaining ten coefficients.
The resulting system of  equations turns out to be
$\xi_j$-independent and linear in the
four variables $\alpha_1$, $\beta_1$, $\gamma_1$ and $\gamma_2$. Therefore we may write the remaining  four equations as
a matrix equation  with coefficients nonlinearly depending on
$\alpha_2$, $\beta_2$, $a_1$ and $a_2$. The rank of the matrix is three.
Solutions (\ref{L1}--\ref{L6}) are obtained by requiring that the matrix be of rank 3, 2, 1  and 0.  In this way we get
\begin{itemize}
\item Case 1:
\beq \label{S1}
\left\{ \begin{array}{l}
\alpha_2 = \beta_2 = 0, \\
{\displaystyle{  {\xi_1 = \xi_2}}},\quad{\displaystyle{ {\xi_3 = \xi_4}}}.
\end{array} \right.
\eeq
\item Case 2:
\beq \label{S2}
\left\{ \begin{array}{l}
\alpha_2 =  \beta_2,\quad \alpha_1 = \beta_1,  \\
a_1 =2 a_2 , \\
\gamma_1 = 2 \gamma_2,\\
a_1 (\xi_1 - \xi_2) = - a_1 (\xi_3 - \xi_4) = -2\alpha_2 \gamma_2.
\end{array} \right.
\eeq
\item Case 3:
\beq \label{S3}
\left\{ \begin{array}{l}
\alpha_2 = - \beta_2,\quad \alpha_1 = \beta_1,  \\
a_2 =2 a_1 , \\
\gamma_2 = 2 \gamma_1,\\
a_1 (\xi_1 - \xi_2) = a_1 (\xi_3 - \xi_4)=
- \alpha_2 \gamma_1.
\end{array} \right.
\eeq
\item Case 4:
\beq \label{S4}
\left\{ \begin{array}{l}
\alpha_1 = \beta_1=\frac12 (1+a_1/a_2)\gamma_2, \\
a_2\gamma_1=a_1\gamma_2,\\
a_1 (\xi_1 - \xi_2) = -\alpha_2 \gamma_1,  \\
a_1 (\xi_3 - \xi_4) = \beta_2 \gamma_1 .
\end{array} \right.
\eeq
\item Case 5:
\beq \label{S5}
\left\{ \begin{array}{l}
(a_2-a_1)\beta_2= (a_2+a_1) \alpha_2,  \\
 2 a_1 a_2 (a_1 - a_2) \alpha_1 = (a_1 + a_2) (\gamma_2 a_1^2-\gamma_1
a_2^2),    \\
2 a_1 a_2 \beta_1 = \gamma_1 a_2^2 + \gamma_2 a_1^2,
 \\
(a_2- a_1)(\xi_1 - \xi_2)=
(\gamma_1 - \gamma_2)\alpha_2,  \\
(a_2 - a_1)^2(\xi_3 - \xi_4)= \left[ \gamma_2 (a_2 - 3 a_1 )-
\gamma_1 (a_1 - 3 a_2 )\right] \alpha_2  .
\end{array} \right.
\eeq
\item Case 6:
\beq \label{S6}
\left\{ \begin{array}{l}
(a_2+a_1)\beta_2= (a_2- a_1) \alpha_2,  \\
2 a_1 a_2  \alpha_1 = \gamma_1 a_2^2 + \gamma_2 a_1^2,
 \\
 2 a_1 a_2 (a_1 - a_2) \beta_1 =(a_1 + a_2) (\gamma_2 a_1^2-\gamma_1
a_2^2),    \\
(a_2^2 - a_1^2)(\xi_1 - \xi_2)= \left[ \gamma_1 (a_1 - 3 a_2 )-
\gamma_2 (a_2 - 3 a_1 )\right] \alpha_2,    \\
(a_1+ a_2)(\xi_3 - \xi_4)= (\gamma_2 - \gamma_1)\alpha_2.
\end{array} \right.
\eeq
\end{itemize}

Then we impose the remaining five equations $\CR_1^{(i)}=0$, $0\leq i \leq 4$ to each of the six obtained solutions (\ref{S1}--\ref{S6}). By a straightforward computer aided computation we conclude that Case 1 has four subcases, Cases (\ref{L1}--\ref{L4}),  that pass those conditions while Case 4 has two subcases, (\ref{L5}, \ref{L6}). All the other cases do not satisfy the equations $\CR_1^{(i)}=0$, $0\leq i \leq 4$.  Cases 1 and 4 represent the necessary and sufficient conditions for the coefficients $a$ and $b$ in (\ref{Abruzzo1}) to be real.
A direct calculation proves invariance of the resulting equations with respect to the reduced M\"obius transformation. Let us stress again that Cases (\ref{L1}--\ref{L6}) are $A_{2}$ $C-$integrable.
\endpf
As a consequence of this proposition we can state the following obvious but important Corollary:
\begin{cor}
If the coefficients $a_1,a_2 ,\alpha_1,\alpha_2,\beta_1,\beta_2,$ $\gamma_1,\gamma_2,$
$\xi_1,...,\xi_4$ of equations~$\CQ_+$ do not satisfy the conditions in
Eqs.~(\ref{L1}--\ref{L6}) then $\CQ_+$ is not linearizable.
\end{cor}

Note that the trivial linearizability condition
$\alpha_1=\alpha_2=\beta_1=\beta_2=\gamma_1=\gamma_2=
\xi_1=\xi_2=\xi_3=\xi_4=0$ is contained in eqs.~(\ref{L1}--\ref{L6}) (cases L1 to L6).

A particularly interesting case is when we consider an equation with at most quadratic  nonlinearity. In this case we get the following Proposition:

\begin{prop} \label{the1bis}
If $\xi_1=\xi_2=\xi_3=\xi_4=0$ in equations $\CQ_+$ then the lowest
linearizability conditions are:
\begin{itemize}
\item Cases QL1a and QL1b: $\alpha_2=\beta_2=0$, $\alpha
   _1= \beta _1$, $2 a_1^2-7 a_2 a_1+2 a_2^2=0\Leftrightarrow a_1=
   \frac{1}{4}
   \left(7\pm\sqrt{33}\right)
   a_2$, $\gamma_1=
   \left(1\mp\sqrt{\frac{3}{11}}
   \right) \beta _1$, $\gamma_2=
   \left(1\pm\sqrt{\frac{3}{11}}
   \right) \beta _1$.
\item Cases QL2a and QL2b: $\alpha_2=\beta_2=0$, $\alpha
   _1= \beta _1$, $4 a_1^4-12 a_2 a_1^3+9 a_2^2
   a_1^2-12 a_2^3 a_1+4 a_2^4=0\Leftrightarrow a_1=
   \left(\frac{3}{4}+\frac{1}{
   \sqrt{2}}\pm\frac{1}{2}
   \sqrt{\frac{1}{4}+3
   \sqrt{2}}\right)
   a_2$,
   $\gamma_1= \mp 2
   \sqrt{\frac{1}{41}
   \left(19+18
   \sqrt{2}\right)} \beta
   _1$, $\gamma_2= \pm 2
   \sqrt{\frac{1}{41}
   \left(19+18
   \sqrt{2}\right)} \beta
   _1$.
\end{itemize}
\end{prop}
Cases~QL1a, QL1b are in~case~L1 and cases~QL2a, QL2b are in~case~L2.

Let us consider now the higher order terms of the expansion for the cases L1--L6. Imposing the $A_{3}$ $C-$integrability conditions (\ref{Aurunci2}) on the real and imaginary parts of the coefficients $\tau_{j}$, $j=1,\ldots, 12$, defined in the expression (\ref{Lazio4}), one obtains that the most general $A_{3}$ $C-$integrable equation is represented by:
\begin{equation}\label{cc}
\begin{gathered}
\alpha_{1}=\beta_{1}=\frac{\left(a_{1}+a_{2}\right)\gamma_{1}} {2a_{1}},\qquad \alpha_{2}=\beta_{2}=0,\qquad \gamma_{2}=\frac{a_{2}\gamma_{1}} {a_{1}},\\
\xi_{1}=\xi_{2}=\frac{3\left(a_{1}+a_{2}\right)\gamma_{1}^2} {8a_{1}^2},\qquad \xi_{3}=\xi_{4}=\frac{\left(a_{1}-a_{2}\right)\gamma_{1}^2} {8a_{1}^2}.
\end{gathered}
\end{equation}
The case (\ref{cc}) is the intersection of the cases~(\ref{L1}) and~(\ref{L2}). 
As a consequence of the result (\ref{cc}) we have that an equation of the form $\CQ_+$ which satisfies the Proposition \ref{the1bis}, i.e. dispersive with at most quadratic nonlinearity,  will never be $C-$integrable. 

Moreover it is easy to prove the following theorem: 

\begin{teor}The equations $\CQ_+$ satisfying the $A_3$ $C-$integrability (\ref{cc}) can be linearized by a real M\"obius transformation
$$ u_{n,m}=\frac{\alpha v_{n,m}+\beta} {\gamma v_{n,m}+\delta}
$$
if and only if $\zeta=\displaystyle\frac{\left(a_{1}+a_{2}\right)\gamma_{1}^3} {4a_{1}^3}$. The coefficients of the linearizing transformation are
$$ \beta=0,\qquad \gamma=-\frac{\alpha\gamma_{1}} {2a_{1}},
$$
while the resulting linearized equation is
\begin{equation}\label{eqLin}
v_{n,m}+v_{n+1,m+1}+\frac{a_{2}} {a_{1}}\left(v_{n+1,m}+v_{n,m+1}\right)=0.
\end{equation}
\end{teor}
Equation~(\ref{eqLin}) is the most general linear dispersive equation defined on the square. As the M\"obius transformation used is restricted M\"obius transformation, (\ref{eqLin}) can be assumed to be the canonical equation for this case.

\section{Conclusions}

In this work we presented the complete classification of the linerizable dispersive partial difference equations belonging to the $\CQ_+$ class using multiple scales expansions around a periodic discrete wave solution of the linearized equation, up to the fifth order in the perturbation expansion parameter. The resulting linearizable system depends on three free parameters, and requesting the resulting equation  to explicitly linearize via a M\"obius transformation reduces the free parameters to one.  It is still an open problem the proof of the linearizability of the equation $\CQ_+$ satisfying just the $A_3$ $C-$integrability conditions (\ref{cc}). Maybe by going higher in the perturbation expansion we could be able to fix the parameter $\zeta$ according to the  theorem of last Section.

This calculation shows that the multiple scale expansion can be effectively used to classify discrete equations. 

Work is in progress for the derivation of the integrable class of dispersive partial difference equations belonging to the $\CQ_+$ class. They are provided by a reduced set of equations with respect to the one considered in this case, as $\rho_2 \ne 0$ is just a real constant. Moreover the integrability conditions are given by a larger number of nonlinear algebraic equations for the parameters entering into the equation and, as such, are much harder to solve.

Work is also in progress in the study of the $\CQ_-$ case. In this case we get at the lowest order in the perturbative parameter a nonlinear system of partial differential equations relating the zeroth and the fundamental harmonic. The solution of this equation is the key ingredient in the classification of this class of equations which contains all dispersive equations belonging to the ABS classification of multilinear equations on the square.

%%%%%%%%%%%%%%%%%%%%%%%%%%%%%%%%%%%%%%%%%%%%
%%%%%%%%%%%%%%%%%%%%%%%%%%%%%%%%%%%%%%%%%%%%
%%%%%%%%%%%%%%%%%%%%%%%%%%%%%%%%%%%%%%%%%%%%
\section*{Acknowledgments}

%%%%%%%%%%%%%%%%%%%%%%%%%%%%%%%%%%%%%%%%%%%%
%%%%%%%%%%%%%%%%%%%%%%%%%%%%%%%%%%%%%%%%%%%%
LD and SC have been partly supported by the Italian Ministry of Education and Research, PRIN
``Nonlinear waves: integrable fine dimensional reductions and discretizations" from 2007
to 2009 and PRIN ``Continuous and discrete nonlinear integrable evolutions: from water
waves to symplectic maps" from 2010. RHH thanks the INFN, Sezione Roma Tre and the UPM for their support during his visits to Rome.

%%%%%%%%%%%%%%%%%%%%%%%%%%%%%%%%%%%%%%%%%%%%
%%%%%%%%%%%%%%%%%%%%%%%%%%%%%%%%%%%%%%%%%%%%
%\end{acknowledgments}
%%%%%%%%%%%%%%%%%%%%%%%%%%%%%%%%%%%%%%%%%%%%

%%%%%%%%%%%%%%%%%%%%%%%%%%%%%%%%%%%%%%%%%%%%
%%%%%%%%%%%%%%%%%%%%%%%%%%%%%%%%%%%%%%%%%%%%

\end{document}